\documentclass[useAMS,usenatbib]{mn2e}
\usepackage{graphicx}
\usepackage{color}
\usepackage{soul} 
\usepackage{url}
\usepackage{multirow} 
\usepackage{txfonts}
\usepackage{longtable}

\usepackage{amssymb}

\definecolor{myred}{rgb}{0.7,0.0,0.2}

\definecolor{myblue}{rgb}{0.0,0.2,0.7}

\definecolor{mygreen}{rgb}{0.2,0.7,0.0}

\title[LMC S63: a deeply eclipsing symbiotic star]{ LMC S63: a historical reappraisal of the outburst behaviour \\
of a deeply eclipsing Magellanic symbiotic star\thanks{Based on observations made with the Southern African Large Telescope (SALT) under programme 2012-2-RSA\_POL\_001}}
\author[I{\l}kiewicz et al.]{Krystian I{\l}kiewicz,$^{1,2}$\thanks{E-mail: ilkiewicz@camk.edu.pl} Joanna Miko{\l}ajewska,$^{1}$ Brent Miszalski,$^{3,4}$
\newauthor Mariusz Gromadzki$^{5,6}$ and Patricia A. Whitelock$^{3}$ \\
$^{1}$Nicolaus Copernicus Astronomical Centre, Bartycka 18, 00716 Warsaw, Poland \\
$^{2}$Warsaw University Observatory, Al. Ujazdowskie 4, 00-478 Warszawa, Poland \\
$^{3}$South African Astronomical Observatory, PO Box 9, Observatory, 7935, South Africa \\
$^{4}$Southern African Large Telescope Foundation, PO Box 9, Observatory, 7935, South Africa \\ 
$^{5}$ Millennium Institute of Astrophysics, MAS, Santiago, Chile\\
$^{6}$ Instituto de F{\'i}sica y Astronom{\'i}a, Universidad de Valpara{\'i}so, Av. Gran Breta\~{n}a 1111, Playa Ancha, Casilla, 5030, Chile
Playa Ancha, Casilla 5030, Chile }

\begin{document}

\date{Accepted xxx. Received xxx; in original form xxx}

\pagerange{\pageref{firstpage}--\pageref{lastpage}} \pubyear{0000}

\maketitle

\label{firstpage}

\begin{abstract}
We present an analysis of multi-epoch low-resolution spectrophotometry, complemented by the light curves provided by massive photometric surveys spanning over 100 years, 
of the symbiotic binary LMC S63. We showed that it is an eclipsing binary with the orbital period of 1050$^{\rm d}$. We  also found evidence of outbursts in history of the white dwarf. If it was a Z-And type outburst, as is most likely, it would  be a second such outburst recorded in the Magellanic Cloud symbiotic system. We confirmed that the red giant is enhanced in carbon, and estimated C/O $\simeq$ 1.2 by fitting a model atmosphere to the SALT spectrum. We also found bi-periodic pulsations of the red giant, and demonstrated that it is similar to other carbon variables with confirmed bi-periodicity.
\end{abstract}


\section{Introduction}
Symbiotic stars  are close binaries consisting of a late-type giant transferring material via wind or Roche-Lobe overflow to a much hotter compact companion surrounded by an ionized nebula. In case of a Mira as the red component a warm dust shell is present in the system (D-type symbiotic) in contrast to S-type with a normal giant (see e.g. \citealt{mikolaj12} for details).

Currently there are 8 confirmed symbiotic stars in the LMC and 8 in the SMC (\citealt{belcz00}; \citealt{oliveira13}; Miszalski, Miko\l{}ajewska \& Udalski 2014). They are amongst the hottest and brightest discovered symbiotics \citep{mikolaj04}.
Additionally all of them contain asymptotic giant branch (AGB) stars, while both red giant branch (RGB) and AGB stars are found in the Galactic symbiotic systems \citep{mikolaj04}. This could be caused by a selection effect and the situation could change as more results from ongoing surveys are published (e.g. Miszalski, Miko\l{}ajewska \& Udalski 2013; \citealt{mm14}; \citealt{rf14}).

Early observations of LMC~S63 showed $\sim$2 mag variability with a period of $\sim$1020 days exhibiting irregularities around maximum (LMV~1485 light curve in \citealt{gapo70})\footnote{Note that the period of 1200 days quoted by \citet{feast74} is a printing error.}. 
Based on this early light curve, \citet{payne71} classified LMC~S63 as an R~CrB variable. First spectra obtained by \citet{feast74} showed  a~smooth continuum  with high excitation emission lines, in particular \mbox{H\,{\sc i}}, \mbox{He\,{\sc i}}, \mbox{[O\,{\sc iii}]} and broadened \mbox{He\,{\sc ii}},
as expected for a~symbiotic star. On further observations a~late-type star continuum was present confirming the symbiotic nature of the system \citep{allen80}. \citet{allen80} suggested that a carbon star could be the red giant component of LMC~S63, which was later confirmed by \citet{morgan92}. Orbitally related variability in photometric observations with a period of 1060$^d$ was found by \citet{mikolaj04}.

To study this star we carried out multi-epoch low-resolution spectrophotometry. The spectra were complemented by the light curves provided by massive photometric surveys.
In Sect.~\ref{data} we describe the data and their reductions. The photometric and spectroscopic variability is analysed in Sect.~\ref{variability}.
The physical parameters of the binary component are derived and discussed in Sect.~\ref{parameters}. A brief summary of our results
is given in Sect.~\ref{summary}.

\begin{figure*}\centering
\resizebox{\hsize}{!}{\includegraphics{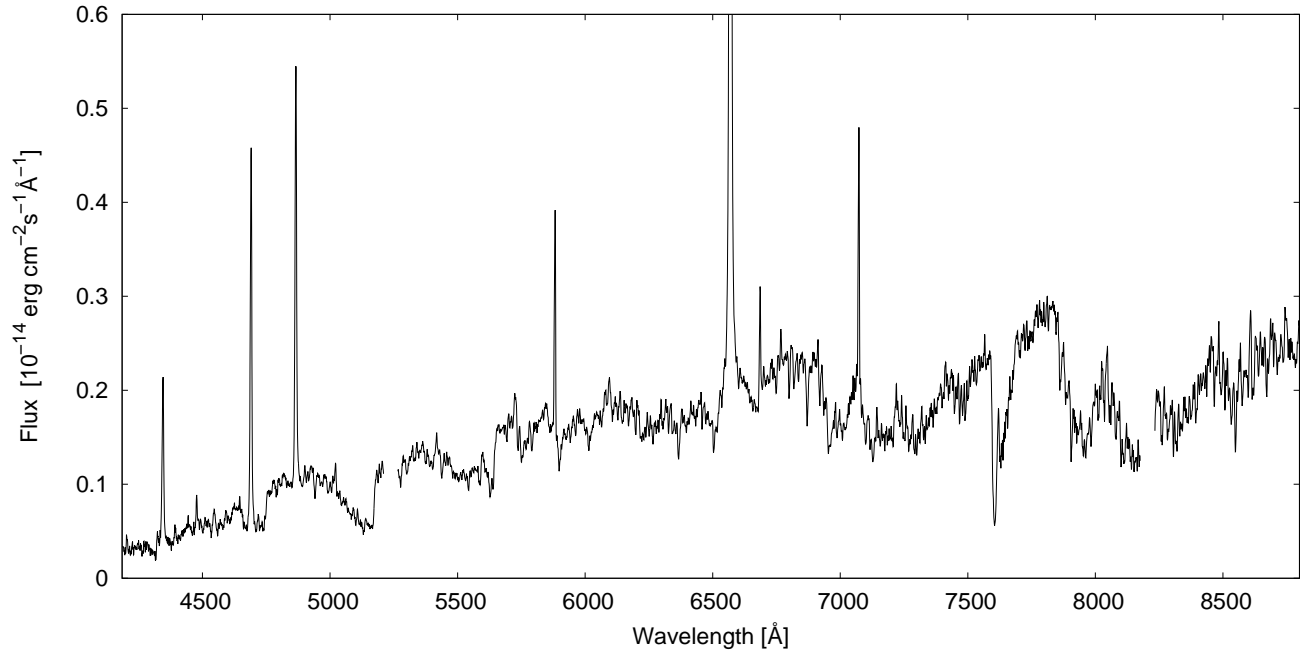}}
  \caption{The SALT RSS spectrum of LMC~S63.}
\label{salt_spectrum}
\end{figure*}

\section{Observations}\label{data}

\subsection{Spectroscopy}

Our spectroscopic observations are summarized in Table~\ref{tablelog}.
Low resolution spectra were made with Grating Spectrograph with the SITe (Scientific Imaging Technologies, INC.) CCD on the 1.9m  telescope  at the South African Astronomical Observatory (SAAO) during the period from 2005 to 2008. 
A long slit with a projected width of 1.5'', and grating \#7 with 300 lines mm$^{-1}$  were used which resulted in a~resolving power of R$\approx$1000. Every night at least one standard star was observed allowing for spectrophotometric calibration. Wavelength calibration was performed using CuAr reference lamp spectra. The covered spectral range varied with every spectrum, but was approximately equal to 4000-7000~\AA.  The standard IRAF procedures were used to reduce the CCD observations and to perform flux calibration. Unfortunately,  due to various reasons (like changes in sky conditions, slit losses, seeing) the~flux calibration was affected by a~gray shift. Therefore we scaled the spectra using known $V$ magnitudes, and the cool giant features (see sec.~\ref{giant}) 

\begin{table}
 \centering
  \caption{The log of observations. The magnitudes are mean magnitudes predicted from the MACHO and OGLE $V$ light curve (Fig.~\ref{phaseall}).}\label{tablelog}
  \begin{tabular}{|cccccc|}
  \hline
Date & MJD & Phase & Telescope & Exposures [s] & V  \\
  \hline
14.11.2005	& 53690 & 0.62 & SAAO 1.9m & 2x1000, 1500 & 15.8\\
14.10.2006	& 54022 & 0.94 & SAAO 1.9m & 3x1000 &  16.2\\
21.10.2007	& 54394 & 0.29 & SAAO 1.9m & 3x1200 &  15.8\\
06.11.2008	& 54776 & 0.66 & SAAO 1.9m & 3x1200 &  15.8\\
18.11.2012  & 56249 & 0.06 & SALT & 2x1800& 16.0\\
05.02.2013  & 56328 & 0.14 & VLT  & 6x440 (UVB) & 15.9 \\
&   &   &    &  6x350 (VIS)  & \\
&   &   &   & 16x180 (NIR)  &   \\	
\hline
\end{tabular}
\end{table}

Another low resolution spectrum was obtained with the Robert Stobie Spectrograph (RSS; \citealt{bur03}; \citealt{kobu03}) on the
queue-scheduled Southern African Large Telescope (SALT; \citealt{buc06}; \citealt{odon06}) under programme
2012-2-RSA\_POL-001 (PI: Miszalski). Two 1800 s exposures were taken with the PG900 grating on 2012 November 18 to cover $\lambda\sim$4200--7275 \AA\ and $\lambda\sim$6155--9150 \AA. After basic calibrations were applied with the \textsc{pysalt} package \citep{cra10}, we followed the same reduction process  as outlined in \citet{MMU14}. 
The relative flux-calibrated spectrum was then normalized  before scaling the spectrum to $V=16.0$ mag, and the resultant spectrum is shown in  Fig.~\ref{salt_spectrum}. The emission line fluxes are listed in Table~\ref{table_flx}.

In addition, we used a publicly available spectrum obtained with X-shooter \citep{vern11} on the UT2 Very Large Telescope (VLT) at the European Southern Observatory under programme  090.D-0422(A). The spectrum was made simultaneously in three arms (UVB, VIS and NIR) which together covered the spectral range 3000-24000\AA. The slit widths for UVB, VIS and NIR arms were 1.0'', 0.9'' and 0.9'', and resulting spectral resolutions R=4350, 7450 and 5300, respectively. 

We supplemented our own spectroscopy with 2 previously published spectra. The first one was obtained by D.~Allen, in 1978, and analysed by \cite{pereira95}. 
The second one is from the spectrophotometric atlas of symbiotic stars (\citealt{munar02}, see therein for details).
The emission line fluxes measured for these spectra are also included in   Table~\ref{table_flx}. 
In addition, we collected emission line fluxes published in \citet{kafat83}, \citet{morgan92},Van Winckel, Duerbeck \& Schwarz (1993) and Muerset, Schild \& Vogel (1996).

We measured emission lines in the UV on three International Ultraviolet Explorer (IUE) spectra and one HST Faint Object Spectrograph (FOS) spectrum. Fluxes derived from three spectra were already presented in the literature (\citet{vogel95} - FOS spectrum, \citet{kafat83} - swp16591, \citet{pereira95} - swp25791), nevertheless since the flux calibration has been improved over the years (see eg. Rodr\'{i}guez-Pascual, Gonz\'{a}lez-Riestra \& Schartel 1999) we repeated the measurements. The results are presented in Table~\ref{table_uv}. 

A typical error of the measured fluxes is 15\% for the strongest and 30\% for the weakest lines and lines in the UV range. This mostly represents uncertainty in selection of the level of the continuum and flux calibration. The collected, previously published fluxes, had similar uncertainty.

\subsection{Photometry}

\begin{figure*}\centering
\resizebox{\hsize}{!}{\includegraphics{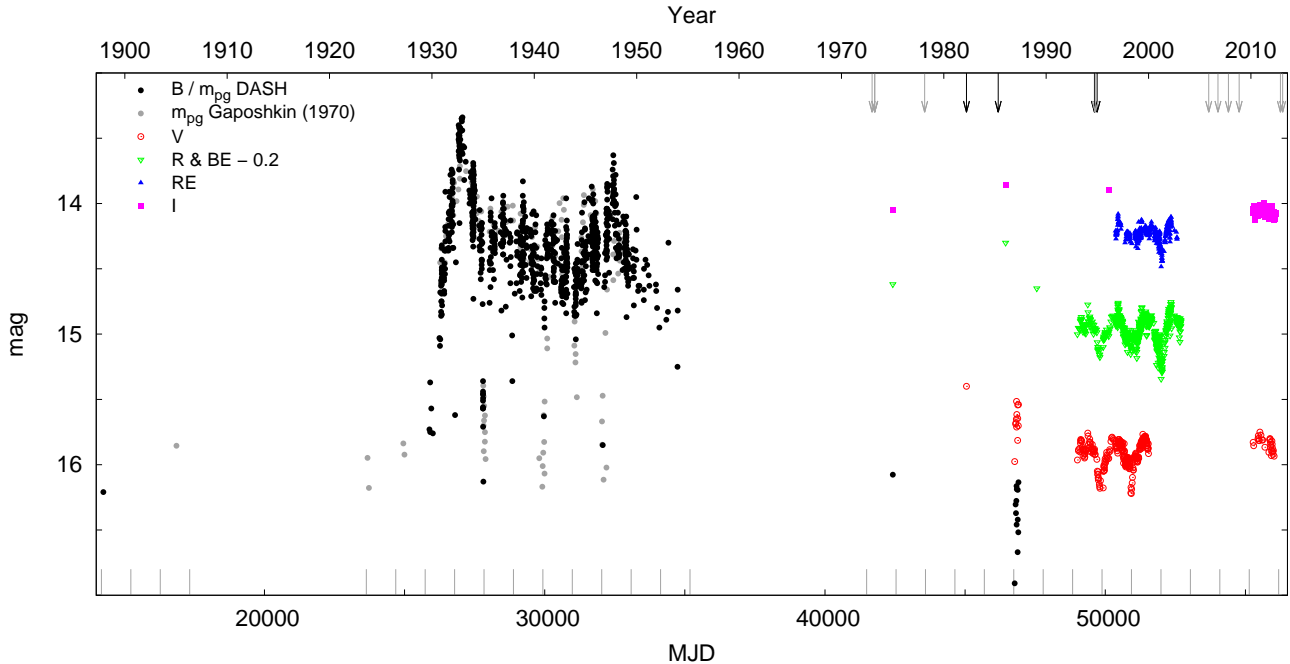}}
  \caption{The DASCH and \citet{gapo70} (black and gray circles, respectively) light curve of LMC~S63. The variability is caused by the eclipses (see text). The vertical lines indicate eclipse times according to our ephemeris. The arrows indicate the dates of spectroscopic observations discussed in the paper.}\label{phot_mjd}
\end{figure*}

The photometric data in this study are from the OGLE (\citealt{udalski08}; \citealt{soszynski11}), MACHO \citep{alcock91} and EROS-2 \citep{spano11} projects. The typical error of measured magnitudes in these surveys was 0.002 mag, 0.003 mag and 0.05 mag respectively.
The OGLE photometry is reduced to the Cousins $V$ and $I$ system. 
The MACHO data were transformed to the Cousins $V$ and $R$ using formulae from \citet{Lutz2010}.  The colour applied for transformation was a colour measured during  individual nights.
The EROS-2 data were taken simultaneously in the non-standard blue BE
(420--720 nm) and red RE (620--920 nm) bands. 
The Harvard College Observatory archival plates covering the field including  LMC~S63, the same as those used by \cite{gapo70}, are now digitalized within the framework of the DASCH  (Digital Access to a Sky Century @ Harvard, \citealt{lay10}) project.  However, a comparison between the DASCH light curve and that published by \citet{gapo70} shows that some original plates must had been lost or perhaps not scanned. Therefore, we scanned Gaposhkin's original light curve and scaled it to fit the DASCH $B$ magnitudes from the same date. Error of magnitude estimates from photographic plates varied between 0.05 and 0.20 mag, which did not include uncertainty in the calibration process.

Single photometric measurements for LMC~S63 are also reported in the literature. These measurements are collected in Table~\ref{table_phot}.


\begin{figure*}\centering
\resizebox{\hsize}{!}{\includegraphics{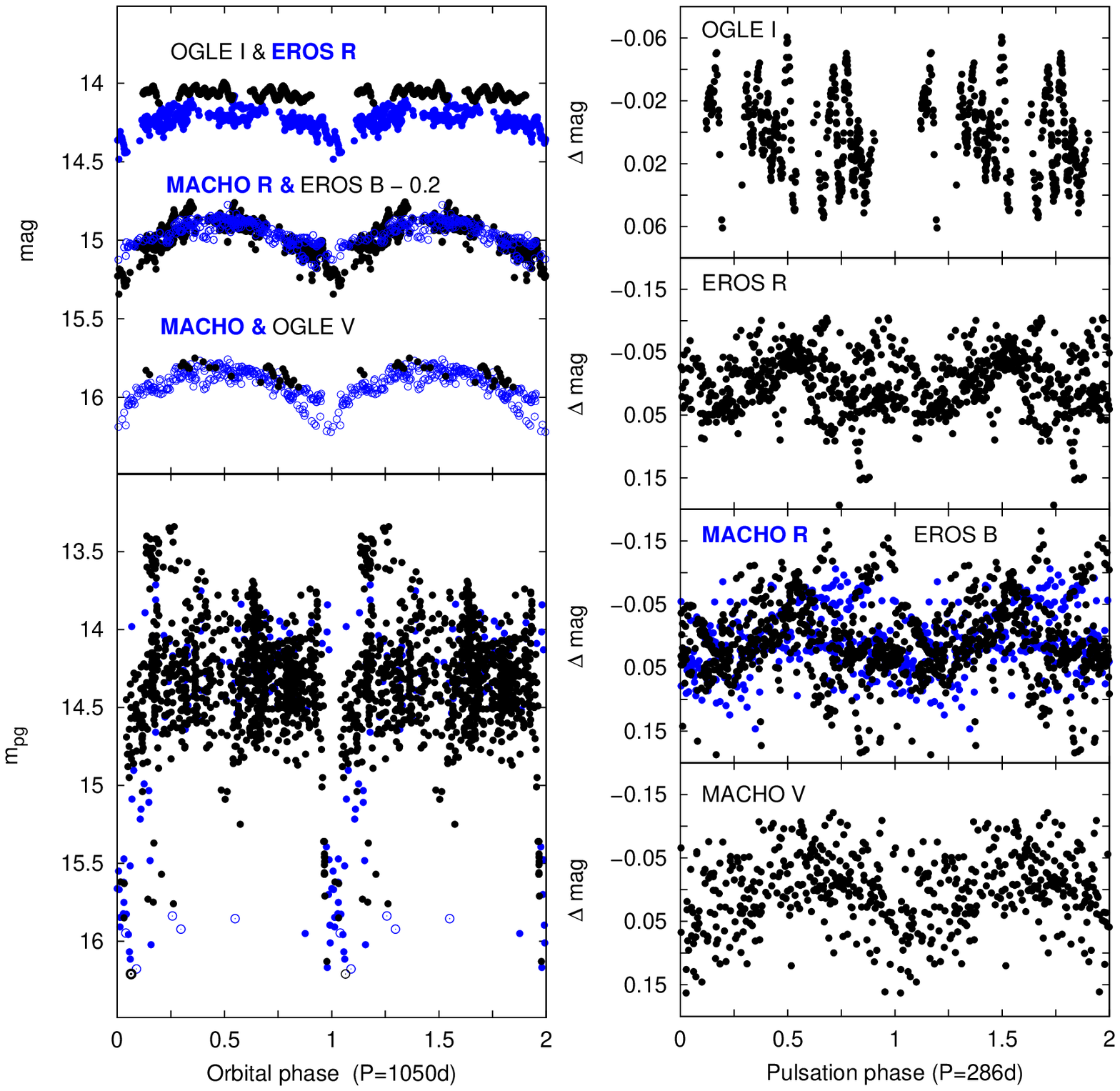}}
 \caption{ Left: Phase plot of the MACHO V \& R data, OGLE V \& I data (black and gray circles respectively), and EROS B \& R data with the orbital ephemeris. Bottom panel: phase plot of the m$_{pg}$ photometric data from \citet{gapo70} and DASCH (gray and black circles, respectively) before and after MJD=25000 (open and filled circles respectively). 
Right: Phase plot with P=286$^d$ of MACHO V \& R data and OGLE I data after subtraction of orbitally related variability.}\label{phaseall}
\end{figure*}

\section{Variability}\label{variability}

\subsection{Photometric variability}\label{phot_var}

The combined light curve of LMC~S63 is presented in Fig.~\ref{phot_mjd}.  
This light curve is typical of eclipsing symbiotic stars which show deep and narrow eclipses during active phases, and more or less sinusoidal light changes during quiescence (e.g. fig. 2 in \citealt{mikolaj03}). 

The deep minima in the  DASCH light curve  can be reproduced with an~ephemeris $MJD_{MIN}=(27836.4\pm7.2) + (1049.7\pm3.5)\times E$. 
Combining this ephemeris with the~three minima visible in the~MACHO and EROS light curves we found the orbital ephemeris to be 
$$MJD_{MIN}=(49885.6\pm7.4)+(1050.0\pm0.5)\times E$$
in agreement with the~1060$^d$ orbital period estimated by \citet{mikolaj04} from MACHO light curves. 
Two of the eclipses in the period covered by the DASCH photometry are not detected; they were most probably missed due to their relatively short duration. 
The first Harvard observations were obtained in 1897 (DASCH) and 1905 \citep{gapo70}, respectively, and in both LMC S63 has photographic magnitude $m_{\rm pg} \sim 16$. Whereas the DASCH point was taken near minimum ($\phi=0.07$),   Gaposhkin's point is from the orbital phase of $\sim$0.5, which strongly suggests that at that time the star was not yet in the~active phase.

In comparison to the  most recent photometry (Fig.~\ref{phaseall}) the minima in the DASCH/Gaposhkin light curve
are very narrow and have a much larger amplitude. This suggests that LMC~S63 was in an active phase in 1930-54 when the DASCH/Gaposhkin  light curve was obtained. 

Combining the~OGLE, MACHO, and EROS data it is clear that the depth of minima during quiescence are larger in bluer filters. This confirms that the variability is related to the hot component obscuration (Fig.~\ref{phaseall}).

The OGLE light curves were also analysed by \citet{angel13}, who found low amplitude changes with a 72-day period 
superposed on a general, slowly decreasing, trend. They attributed these periodic changes to red giant pulsations, and classified it as a LMC OSARG. However, they did not discuss the long-term trend because their baseline is too short. 
We can reveal with our longer time coverage 
that this decreasing trend is due to the orbital variability. 

A power spectrum of the MACHO data after subtracting a~sinusoid with a period of $1050^{\rm d}$ shows the most prominent peak at 286$^d$. Unfortunately the~MACHO data are not adequate for searching for 72$^d$ variations because of the~low time-resolution and the amplitude of the variability is similar to the accuracy of the measurements. EROS observations cover a period of time which partially overlaps the time period covered by MACHO. The EROS observations confirm the 286$^d$ pulsations (Fig.~\ref{phaseall}). Episodically the 72$^d$ oscillation seems to also be detectable. Presumably during this time observations were carried out during the best observing conditions which allowed for detection of variability with such small amplitude. In the~OGLE data the~286$^d$ modulation is not found (Fig.~\ref{phaseall}). The cause of this could be the fact that the~OGLE data mainly consists of a~separate $\sim$300$^d$ datasets so it is not sufficient for searching for changes with such long periods. We conclude that in the time period covered by MACHO and EROS data (1993-2003) both the 72$^d$ and the 286$^d$ variability was present but the 286$^d$ variability most probably stopped before the OGLE observations (2010-2013). This would be in agreement with the fact that $>200^d$ variations superimposed to the variability on a time scale of tens of days could be episodic in a~small-amplitude red variables, as was observed eg. in W~Boo \citep{perc96}.

The average $K=11.3$ locates LMC S63 above the First Giant Branch (FGB) tip, whereas the 286$^d$ and 72$^d$ periodicities are consistent with the Mira sequence and  the B sequences, respectively, in the $K\, \log P$ plane \citep{wood2000}. 
This  suggests that the longer period may represent the fundamental mode and the shorter one the first or a higher overtone.
Based on studies of nearly 300 Galactic carbon giants, Bergeat, Knapik \& Rutily (2002a) discussed their physical parameters. Using the average $<J-K>_{\rm 0}$=1.14 and $<K>_{0}$=11.3 for LMC S63 (Table A1), and adapting their calibration of the bolometric correction BC(K) vs $J-K$ colour we estimate $M_{\rm bol}$=-4.4.
Adopting the effective temperature $T_{\rm eff}=3500\, \rm K$ consistent with the best fit to the optical spectrum and the IR SED (see section~\ref{giant}) we derive the radius $R_{\rm g}$=184 $R_{\sun}$, which is in very good agreement with the value resulting from our model fit to the optical spectrum, although both values could be affected by a systematic error.
The red giant of LMC S63 then lies on the evolutionary track of a low metallicity (Z=0.008) 1.9 M$_{\sun}$ star in the HR diagram given in Fig. 9 of \citet{bergeat2002a}.

The pulsation properties, and in particular the bi-periodicity of the Galactic carbon-rich variables was studied by Bergeat, Knapik \& Rutily (2002b), who gave relations between their periods and physical parameters.
In particular, the two periodicities found in S63 locate it in the period-radius diagram (Fig. 2 of \citealt{bergeat2002b}) among other carbon variables with confirmed bi-periodicity. However,  its radius and the 286$^{\rm d}$ period is consistent with  the fundamental mode only if the pulsation mass is $\sim 0.8\, \rm M_{\sun}$, much lower than the initial mass given by the position of S63
in the HR diagram. The ratio of the two periods, $P_0/P_1\sim 4$ is much higher than the average value of $\sim 2.24$ found by \citet{bergeat2002b}, and the mean theoretical value, $\ga 2$ for fundamental to first overtone modes.

\begin{figure}
\centering
\resizebox{\hsize}{!}{\includegraphics{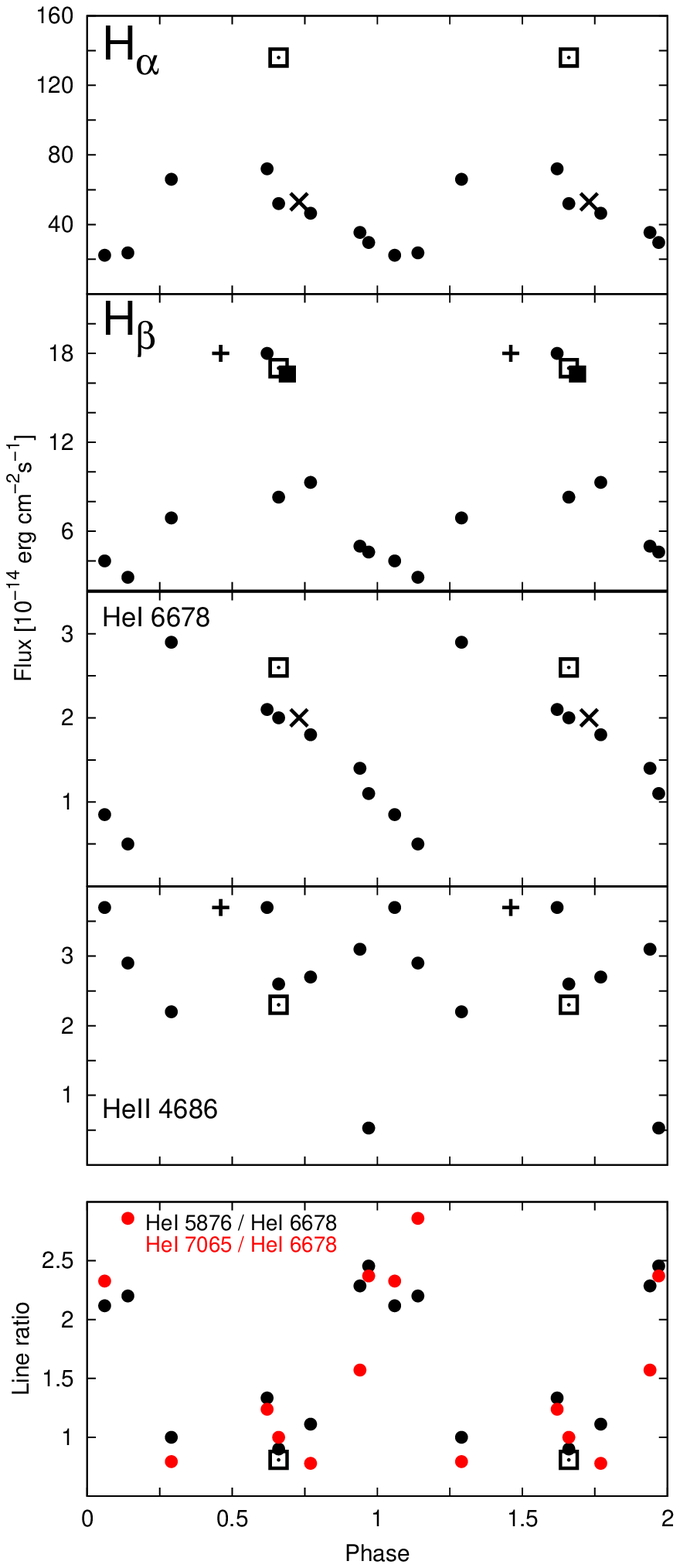}}
  \caption{Phase plot of fluxes of selected emission lines. The fluxes are highest around phase~0.5 as expected for orbital modulation. 
However, the \mbox{He\,{\sc i}} line ratios show a rise near phase 0 (the hot component behind the red giant).
References: $\bullet$ - this study, $\boxdot$ - \citet{kafat83}, $\blacksquare$ - \citet{morgan92}, $\times$ - \citet{win93}, $+$ - \citet{murs96}.}\label{el_phase}
\end{figure}

\subsection{Spectroscopic behaviour}\label{sp_var}

We present measured fluxes of the~emission lines in our spectra in Table~\ref{table_flx}. Phase plots of the~fluxes of selected emission lines are presented in Fig.~\ref{el_phase}. The fluxes of the \mbox{H\,{\sc i}} and \mbox{He\,{\sc i}} lines are lowest during the~photometric minimum. Such behaviour is expected for orbitally related modulation and is caused by partial obscuration of the~line forming region. Similar variability of the~emission lines was observed, e.g., in CI Cyg \citep{kenyon91}, and many other S-type symbiotic stars.
Such orbitally related changes should be also observed in the Balmer UV continuum. In fact, the Galaxy Evolution Explorer (GALEX) (\citealt{martin2005}; \citealt{morrissey2007}) observations of LMC~S63 resulted in F(NUV)=647.59+/-13.48 mJy on MJD 55803 ($\phi=0.700$), and F(NUV)=553.52+/-7.92 mJy on MJD 55903 ($\phi=0.796$), and they seem to reflect mostly the expected changes in the Balmer continuum.

The \mbox{He\,{\sc i}}\,5876/6678 and 7065/6678 line ratios increase near the orbital phase $\phi=0.0$ (hot component obscuration; see Fig.~\ref{el_phase}). The reason for this is unclear but it is worth mentioning that these two lines are more likely enhanced due to a collisional excitation than the \mbox{He\,{\sc i}\,6678} line (see eg.~\citealt{cle87}). The change in ratios of the lines are similar to those expected for changes associated with modulations of electron density in symbiotic binaries (Proga,  Miko\l{}ajewska \& Kenyon 1994). This is consistent with the \citet{allen80} discussion about the fact that the high ratio of the \mbox{He\,{\sc i}} lines to the Balmer lines derived from his spectrum ($\phi=0.95$) indicates unusually high He abundance in the nebular region or collisional excitation of \mbox{He\,{\sc i}}.

\begin{figure}
\resizebox{\hsize}{!}{\includegraphics{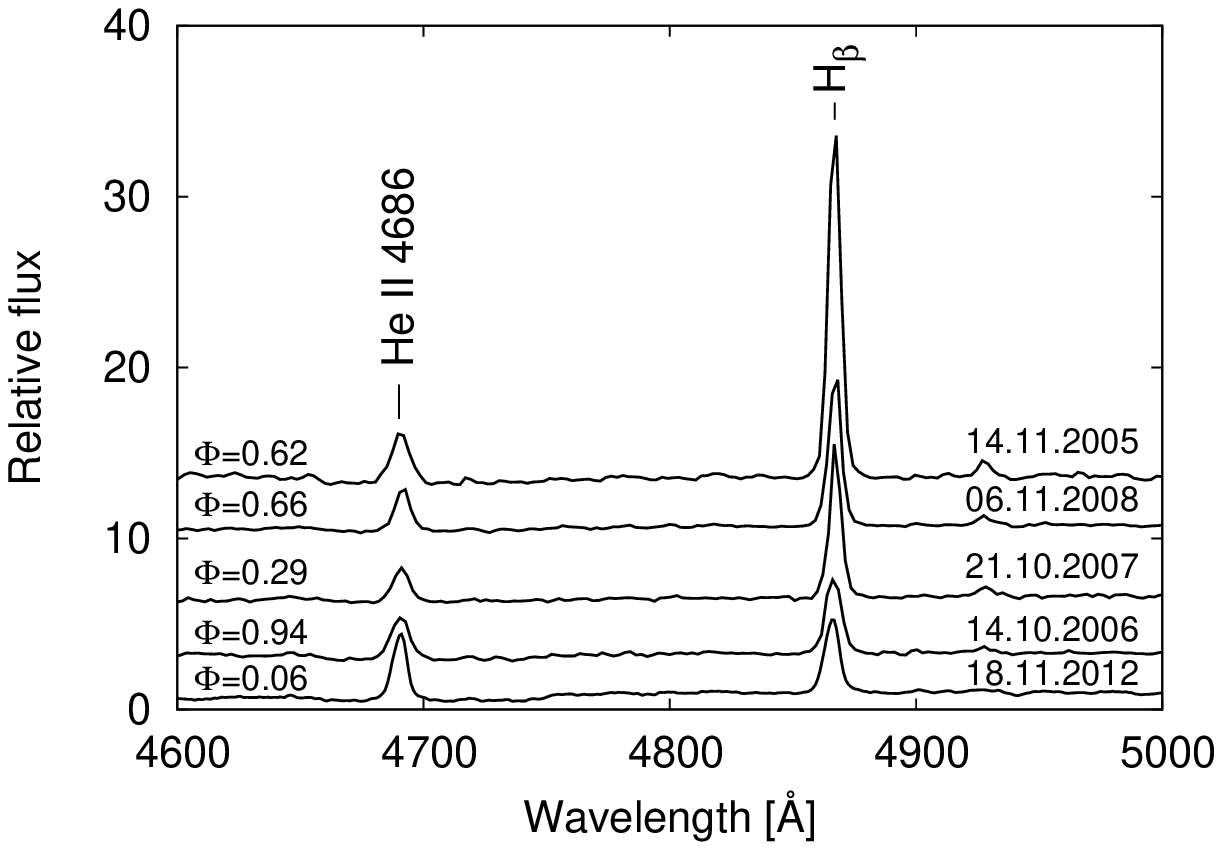}}
  \caption{Comparison of orbitally related variability of \mbox{He\,{\sc ii}\,4686} and H$_\beta$ in the spectra taken in 2005-2012.}\label{heIIhbeta}
\end{figure}

\begin{figure}
\centering
\resizebox{\hsize}{!}{\includegraphics{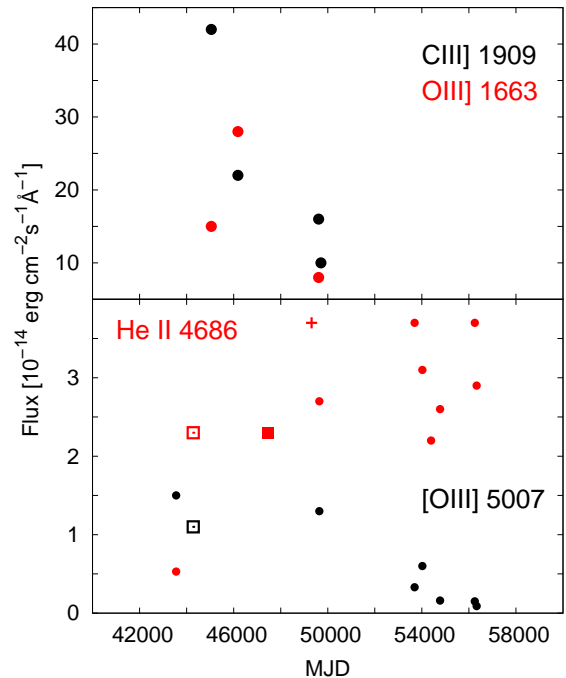}}
  \caption{The gradual decrease of the~\mbox{[O\,{\sc iii}]} and \mbox{C\,{\sc iii}]} line strength and increase of \mbox{He\,{\sc ii}\,4686} line. The~\mbox{O\,{\sc iii}]} line shows gradual decline superimposed to the orbital variability. References: $\bullet$ - this study, $\boxdot$ - \citet{kafat83}, $\blacksquare$ - \citet{morgan92}, $+$ - \citet{murs96}.}\label{oiiispadek}
\end{figure}

\mbox{He\,{\sc ii}\,4686} shows no orbitally related modulations in contrast to the~other lines (Fig.~\ref{heIIhbeta}), except for the spectrum taken during the 1978 eclipse ($\phi=0.97$) when it was only marginally detectable.
It is possible that the eclipses in the \mbox{He\,{\sc ii}} lines are very narrow or the line shows eclipse behaviour depending on its activity stage.
Similar behaviour is observed in many other symbiotic stars, which show deep and narrow eclipses in \mbox{He\,{\sc ii}\,4686} only during some phases of their outbursts and nearly constant flux  in quiescence (e.g. CI~Cyg; \citealt{kenyon91}). Moreover the line shows slow rise over the observed period (Fig.~\ref{oiiispadek}).  Such behaviour can be accounted for by changing conditions and stratification effects in line forming region(s). 

The only forbidden lines present in our spectra are those of \mbox{[O\,{\sc iii}]}. Whereas  \mbox{[O\,{\sc iii}]}~5007 is visible during the whole observed period \mbox{[O\,{\sc iii}]}~4363 was  detected only in 1978, 1997 and 2006. Both lines show a gradual decline (Fig.~\ref{oiiispadek}). 
At the same time \mbox{He\,{\sc ii}\,4686} shows a gradual increase. Similar behaviour was observed in CI Cyg by \citet{kenyon91} during late decline from its major 1975 outburst. 

We recall here that the first spectra of LMC S63 taken by \citet{feast74} in  1972/73 ($\phi \sim  0.2$) revealed a broadened  \mbox{He\,{\sc ii}\,4686} line with FWHM$\simeq$300\,km/s, which suggests formation near to the hot component, presumably in its wind. The line was even broader, with FWHM of 770 km/s, in late 1993 \citep{murs96}. The broadening of the  \mbox{He\,{\sc ii}\,4686} with respect to other nebular lines, including \mbox{H\,{\sc i}}, 
and  \mbox{[O\,{\sc iii}]} was also noticed by \citet{allen80} and \citet{kafat83}.
On the VLT spectrum the  \mbox{He\,{\sc ii}\,4686} line is not so broad, its FWHM is only 134 km/s. 
Broadening of emission lines is a common characteristics of symbiotic outbursts, and it is usually attributed to increased wind and expansion of the hot component atmosphere.
LMC~S63 very likely underwent an optical outburst in the 1930/50s. However, it is hard to say whether the presently observed secular changes are related to that event or another outburst had happened more recently. Nevertheless there was no additional activity of the system recorded in the archive data, although there is a gap from 1954-75 (see Table~\ref{table_phot}). 

\section{Parameters of the components}\label{parameters} 

\subsection{The cool giant}\label{giant}

The cool component is a~carbon star, which was first suggested by \citet{allen80} and confirmed by \citet{morgan92}. 
\citet{murs96} estimated its spectral type as type C2.1J  using the monochromatic flux ratios defined by \citet{Cohen1979} based on the classification scheme proposed by \citet{Yamashita1972}.
However, the only attempts to determine the chemical abundances of this system were carried out for the nebula (\citealt{vogel95}, \citealt{pereira95}) and resulted in a carbon abundance too low for a~carbon enriched star. To determine the carbon abundance in the giant we attempted fitting synthetic spectra to the observed spectrum. Synthetic spectra were calculated using \citet{cas03} models and the SPECTRUM code \citep{Gray94} with help of an updated \citet{plez98} TiO lines list from the authors website\footnote{http://www.pages-perso-bertrand-plez.univ-montp2.fr/}. The adapted solar abundances are from \citet{Grev98}.

The input effective temperature, $\mathrm{T_{eff}}$, was estimated from the relation $T_{\rm eff}=7070/[(J-K)+0.88]$ derived by Bessell, Wood \& Evans (1983) for K, M and C giants, for which $J-K=1.17$ (Table A1) resulted in  $T_\mathrm{eff}=3450 \mathrm{K}$. To compute the synthetic spectrum we also assumed the mean LMC metallicity $\mathrm{[M/H]}=-0.5$, and $\log \mathrm{g}=0$ (appropriate for a low mass star with a radius of $\sim 200\, R_{\sun}$, e.g. \citealt{murs96}). 

\begin{figure}
\centering
\resizebox{\hsize}{!}{\includegraphics{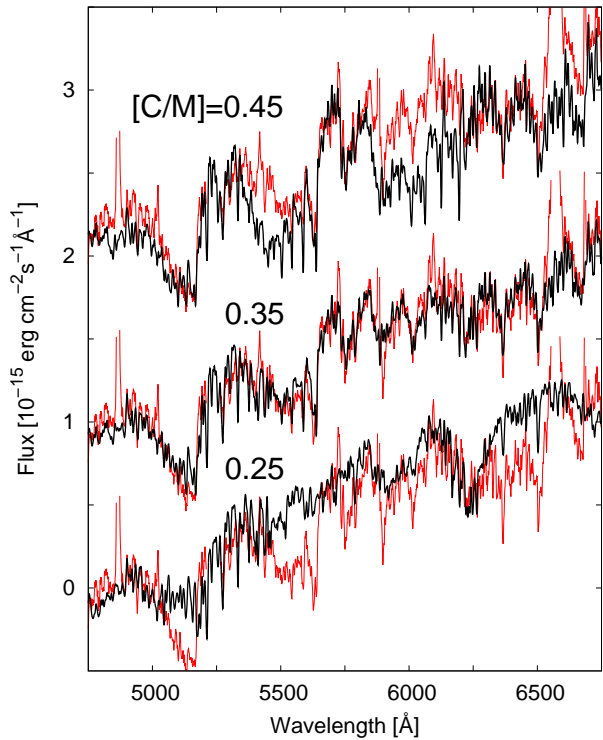}}
  \caption{Scaled model atmosphere spectra (black lines) with increasing carbon abundance 
compared to the SALT spectrum (red lines).}\label{fit_model}
\end{figure}

After calculating a grid of models we performed a least squares fit of a scaled model to the SALT spectrum in the spectral range 5000$\AA$-6500$\AA$. As a result we obtained a best fit for $\mathrm{[C/M]}=0.35\pm0.02\,\mathrm{dex}$ and  $T_\mathrm{eff}=3500\pm250\mathrm{K}$ (see Fig.~\ref{fit_model}).
The scaling factor, $(R_{\rm g}/d)^{-2}=7.48 \times 10^{-21}$, corresponds to the red giant radius $R_{\rm g}=190\, \rm R_{\sun}$ at the distance to the LMC, $d=49.97$ kpc \citep{pietrz13}.

The fit should not be affected by the~nebular emission because the SALT spectrum was taken near to the minimum ($\phi = 0.06$). Moreover, the synthetic photometry of the model fits well to both optical and IR magnitudes confirming that  the nebular contribution is indeed negligible.  
When we assume the same oxygen abundance as the metallicity, and adopting the solar C/O=0.55 \citep{asplund}, we receive  
$\mathrm{C/O} \simeq 1.2$,
a clearly higher value than the one calculated for the nebula by \citet{vogel95}; $\mathrm{C/O}=0.27$ and $\mathrm{C/O}=0.68$ from 1982 and 1994 spectrum, respectively). 
Our result supports their suggestion that
 the C/O ratio in the nebula represents mostly the abundance of the hot component and possibly consists of material ejected during the active phase, which is consistent with the forbidden line variability. The increase of the C/O ratio in the nebula between 1982 and 1994 can be explained by the fact that the material in the nebula is slowly becoming dominated by material from the red giant wind. 

LMC S63 was included in the recent study of SAGE AGB candidates in the LMC by  \citet{riebel12}, who determined $T_{eff}$ of the cold component and the dust mass-loss rate by fitting model to spectral energy distribution based on 12 photometric bands from optical to mid-infared wavelengths. Unfortunately, they misclassified LMC S63 as O-rich star, which could affect their fitting result. We repeated their procedure by fitting the GRAMS C-rich grid of models  of dust shells around red supergiant and AGB stars (Srinivasan, Sargent \& Meixner 2011) to the same mid-infared photometry as  \citet{riebel12}, supplemented with data from WISE \citep{cut12} and DENIS \citep{Cioni00} catalogs. Using VOSA \citep{bayo08} we performed a Bayesian analysis of the data. We assumed the distance of LMC $d=49.97$~kpc \citep{pietrz13} and E(B-V)=0.05 \citep{murs91}. As a result, apart from parameters of the red giant, we estimated such parameters as the dust mass-loss rate  $\dot{M}_{dust}$,  inner radius of the dust shell $R_{in}$ and optical depth at 11.3 microns $\tau_{11.3}$ (see Table~\ref{table_cold_param}). Note that the grid of models available to us had a~solar metallicity and highest possible $\log g=0$, hence this results should be accepted with caution. Particularly the photospheric ratio of C/O$\simeq$2 is  higher than the value resulting from the optical spectra fitting, and it could be affected by the fact that LMC~S63 has lower oxygen abundance than the one used in models.

\begin{table}
 \centering
  \caption{Parameters of the cold component from fitting GRAMS C-rich models to IR data. The values from Bayesian analysis are presented.  }\label{table_cold_param}
  \begin{tabular}{cccc}
\cline{2-3}
 &  Best value & Probability \\\hline
 $T_{\rm eff}$ &  3700 & 54.60 \% \\
L &  9120 L$_\odot$ & 54.60\% \\
$\log \mathrm{g}$ & 0 & 100.00  \% \\
Mass &  2 	M$_\odot$ &100.00  \% \\
C/O &  2 & 52.51 \% \\
$\dot{ \mathrm{M}}_{dust}$  & $2.53\times 10^{-11}$ M$_\odot$/yr & 35.75 \% \\
 $\mathrm{R_{in}}$& 3 R$_\mathrm{star}$ & 51.78 \% \\
$\tau_{11.3}$ & 0.006 & 32.30 \% \\
\hline
\end{tabular}
\end{table}

\subsection{The hot component}\label{hc}

The published estimates of the hot component parameters were based on  fitting a blackbody to the UV spectra in addition to the   
UV and optical  \mbox{He\,{\sc ii}}, and H$\beta$ emission line fluxes. 
Given the eclipsing nature of LMC S63, the only reliable estimates are those based on data collected out of the eclipse, and preferably near to phase $\sim 0.5$ when the hot component passes in the front of the red giant.
In particular, \citet{vogel95}
derived the hot component temperature, $T_{\rm h}$, and luminosity, $L_{\rm h}$, by fitting the  \mbox{He\,{\sc ii}\,1640} emission line flux and the adjacent continuum. 
Their values,  $T_{\rm h}$=80--115kK and 80--90\,kK, and $L_{\rm h}=8200$--$11\,500\, \rm L_{\sun}$ and 5000--$6200\, \rm L_{\sun}$ obtained from the 1982 IUE and 1994 IUE/HST data, respectively, are in fairly good agreement with  other estimates (\citealt{kafat83}; \citealt{murs91}; \citealt{me94}) based on the same IUE data. 

The maximum brightness of LMC~S63, observed during the 1930--1950 optical outburst, and corrected for a reddening $E(B-V)=0.05$ (e.g. \citealt{murs91}) consistent with the foreground  reddening towards the LMC, is $m_{\rm pg}\simeq 13.2$. This corresponds to an absolute magnitude $M_{\rm pg}\simeq  -5.3$, which 
assuming that most of the hot component continuum emission is shifted to the optical (a lower limit if not), results in $L_{h,max}=10\,900 L_\odot$. 

In the 2000's the hot component reached quiescence (as indicated by the optical brigthness, and the changes in the emission lines), and we can make use of the emission line fluxes listed in Table~\ref{table_flx}.
In particular, 
the lower limit on the hot component temperature can be set via the relation 
between the highest observed ionization potential (IP) and the hot component temperature, namely $T [\rm K]\sim 10^3\,IP[\rm keV]$, proposed by \citet{murs94}.
On both our SALT and VLT spectra the highest detected IP was 99~keV for Fe$^{+6}$.  
An upper limit for $T_{\rm} \sim 120\rm kK$ was estimated from  
 \mbox{He\,{\sc ii}\,4686}, \mbox{He\,{\sc i}\,5876} and H$\beta$ emission line ratios observed outside of the eclipses (2005, 2007 and 2008 data in Table~\ref{table_flx}) assuming case B recombination and a black-body continuum \citep{iij81}.
Equations  6 and 7 of Miko\l{}ajewska, Acker \& Stenholm (1997)  then give $L(\mbox{He\,{\sc ii}\,4686} \simeq 1300\, \rm L_{\sun}$, and $L(H\beta) \simeq 1000\, \rm L_{\sun}$. The temperature and luminosity of the hot component of LMC~S63  is then similar to those of other queiscent symbiotic stars (e.g. \citealt{murs91}, \citealt{mikolaj97}).

The high temperature, $T \ga 100$ kK, and luminosity, $L \ga 1000\, \rm L_{\sun}$,  indicates that LMC~S63 could be a detectable soft X-ray source.
Indeed, LMC~S63 was detected at the 5$\sigma$ level by the ROSAT PSPC survey in September~7-December~17,~1990 ($\phi=0.35-0.45$; Bickert, Stencel \& Luthardt 1993). The flux corrected for vignetting was 0.0112 counts/sec which is $\sim$140 times less than the flux of AG~Dra observed at quiescence on Nov~27-Dec~7 \citep{bickert93}, at orbital phase $\phi\sim0.7$ and $L_h\sim1000$ (d/2.5 kpc)$^2$ \citep{mikolaj95}. After scaling the AG~Dra flux to the distance of the LMC \citep{pietrz13} its flux should be equal to $\sim$0.00385 counts/sec, $\sim$3 times fainter than LMC~S63. We recall here that the LMC~S63 hot component luminosity was  $\sim 5 \times$ higher in 1994 than that for AG Dra. 

The presence of broad \mbox{He\,{\sc ii}}\,4686 in the spectra taken in 1974-1993, as well as the presence of broad high excitation lines in the HST spectrum taken in 1994 can be accounted for by the presence of a fast wind from the hot component \citep{vogel95}. The weakness of the \mbox{He\,{\sc ii}} line during the 1978 eclipse of the hot component compared to that on the spectrum taken 2 years later by \citet{kafat83} (see Fig.~\ref{oiiispadek}) additionally supports this interpretation.
However, the \mbox{He\,{\sc ii}} is relatively narrow on the VLT spectrum taken in 2013, which indicates that the wind has ceased by this time, and the line does not show eclipses anymore.
The change in the  \mbox{He\,{\sc ii}} characteristics coincides with a decrease of the hot component luminosity by almost one order of magnitude, which suggests that LMC~S63 has recovered to a quiescent state.
Unfortunately the $\sim$ 20-year gap in photometric observations does not allow us to ascertain whether the spectroscopic evolution was due to continued decline from the 1930-54 optical outburst. Nevertheless, we note that the hot component luminosity in 1982 and 1994 was comparable to that derived for the optical maximum.

It is also not easy to deduce the nature of the outburst, whether this was a symbiotic nova or a Z And-type outburst. 
The light curve (Fig.~\ref{phot_mjd}) shows at least two maxima which is more typical for Z And-type outbursts (see, e.g., the light curves of Z And, AR Pav and other Z And-type systems in the AAVSO database). Some  of the Z And-type systems also show very prolonged active phase (e.g. AR Pav has been active for more than a decade) or a series of outbursts (e.g. Z And). The outburst amplitude,
$\Delta m_{\rm pg} \sim 2.5$ is also rather low compared to other symbiotic novae with their typical optical amplitudes exceeding $\sim 5$ mag (e.g. PU Vul -- see fig. 8 of Kato, Miko{\l}ajewska \& Hachisu 2012).
The pre-outburst photographic magnitude, $<M_{\rm pg}> \simeq 16 \pm 0.3$,  is close to the $B$ mag observed in the 1980-90s, which suggests that the object was brighter than expected for a C-rich red giant, and that the hot component and symbiotic nebula considerably contributed to the observed brightness.
All these speak in favour of Z And-type outburst, which would make LMC~S63 the second Z-And-type Magellanic symbiotic star after LIN~9 \citep{mm14}, although we cannot completely 
 exclude a weak nova outburst on a hot white dwarf.  

\section{Conclusions}\label{summary}

We discussed for the first time orbital variability and outburst behaviour based on photometry spanning more then 100 years, for Magellanic symbiotic system LMC S63. We showed that it is an eclipsing system with a orbital period of $P_{\rm orb}=1050.5 \pm 0.5^{\rm d}$. We also found orbitally related changes in the~emission line fluxes. The DASCH photographic light curve revealed a $\sim 2.5$-magnitude outburst which started in 1930. The star had remained active in the optical for at least 14 years. A broadened \mbox{He\,{\sc ii}\,4686} with respect to other nebular lines revealed by the
first spectra taken in the 1970s, as well as all other spectra taken in 1980s and 1990s, suggests that either LMC S63 had undergone another outburst or the first outburst continued, and the system reached quiescence only in the 2000s.  

We demonstrated that the red giant is enhanced in carbon, and derived C/O $\simeq$ 1.2 by fitting a model atmosphere to the SALT spectrum taken near the optical minimum. We also discussed the pulsation characteristics of the red giant, and compared them to other carbon variables. In particular, the two periodicities found locate the giant component of LMC S63 in the period-radius diagram among other carbon giants with confirmed bi-periodicity.

Some of these conclusions results could not have been made by previous studies based on data spanning only a few years (e.g. Mikolajewska 2004, Angeloni et al. 2014), which also emphasises the need to look at data over a very long baseline to fully understand these systems. 
Because of its complex variability, LMC S63 deserves to be monitored in the future. In particular, another outburst would be among the most interesting thing to observe. 

\section*{Acknowledgments}
We are grateful to Ulisse Munari for making spectrum from his atlas available for this study.  We thank Claudio~B.~Pereira for valuable discussion. 
Some of the observations reported in this paper were obtained with the Southern African Large Telescope (SALT). Other observations were made at the South African Astronomical Observatory (SAAO).
This paper utilizes public domain data obtained by the MACHO Project, jointly funded by the US Department of Energy through the University of California, Lawrence Livermore National Laboratory under contract No. W-7405-Eng-48, by the National Science Foundation through the Center for Particle Astrophysics of the University of California under cooperative agreement AST-8809616, and by the Mount Stromlo and Siding Spring Observatory, part of the Australian National University. This research has made use of the SIMBAD database and VizieR catalogue access tool, operated at CDS, Strasbourg, France. This publication makes use of VOSA, developed under the Spanish Virtual Observatory project supported from the Spanish MICINN through grant AyA2011-24052.
This research was partly supported by the Polish National Science Center grant number DEC-2011/01/B/ST9/06145. KI has been also financed by the Polish Ministry of Science and Higher Education Diamond Grant
Programme via grant 0136/DIA/2014/43.  MG acknowledges support from Joined Committee ESO and Government of Chile 2014 and the Ministry for the Economy, Development, and Tourisms Programa Inicativa Cientifica Milenio through grant IC 12009, awarded to The Millennium Institute of Astrophysics (MAS) and Fondecyt Regular No. 1120601.
PAW thanks the National Research Foundation  for a research grant. 


\newpage

\appendix

\section[]{}

\begin{table*}
 \centering
   \caption{The observed wavelength of emission lines from the VLT spectrum and emission line fluxes in all of our spectra. }\label{table_flx}
  \begin{tabular}{||c|ccccccccc|}
\cline{2-10}

 & Date &	15.02.1978 & 17.10.1994 & 14.11.2005 & 14.10.2006 & 21.10.2007 & 06.11.2008 & 18.11.2012 &  05.02.2013 \\
 & MJD &	    43555 & 49643 &	53690 &	54022 &	54394 &	54776 &	56249 & 56328 \\
 & Orbital phase  &  0.97 & 0.77 & 0.62  &  0.94 & 0.29 &	0.66 & 0.06 & 0.14 \\\hline

$\lambda_{obs}$	[\AA] &	ID	&	\multicolumn{8}{ |c| }{Flux [$\rm 10^{-14} erg\, s^{-1} cm^{-2}$]	} \\\hline
3124.6	&	\mbox{O\,{\sc iii}} 3121.7	&	   &    &	   &    &    &	   &	   & 0.23	\\
3135.6	&	\mbox{O\,{\sc iii}} 3132.9	&	   &    &	   &    &    &	   &	   & 0.95	\\
3205.6	&	\mbox{He\,{\sc ii}} 3203.1	&	   &    &	   &    &    &	   &	   & 0.56	\\
3315.5	&	\mbox{O\,{\sc iii}} 3312.3	&	   &    &	   &    &    &	   &	   & 0.17	\\
3343.8	&	\mbox{O\,{\sc iii}} 3340.7	&	   &    &	   &    &    &	   &	   & 0.25	\\
3348.7	&	[\mbox{Ne\,{\sc v}}] 3345.8	&	   &    &	   &    &    &	   &	   & 0.37	\\
3428.7	&	[\mbox{Ne\,{\sc v}}] 3425.9	&	   &    &	   &    &    &	   &	   & 1.00	\\
3431.0	&	\mbox{O\,{\sc iii}} 3428.7	&	   &    &	   &    &    &	   &	   & 0.36	\\
3447.2	&	\mbox{O\,{\sc iii}} 3444.1	&	   &    &	   &    &    &	   &	   & 0.42	\\
3589.6	&	[\mbox{Fe\,{\sc vii}}] 3587.2	&	   &    &	   &    &    &	   &	   & 0.08	\\
3706.6	&	H $_{16}$	&	   &    &	   &    &    &	   &	   & 0.02	\\
3724.8	&	H $_{14}$ 	&	   &    &	   &    &    &	   &	   & 0.03	\\
3737.2	&	H $_{13}$	&	0.10   &    &	   &    &    &	   &	   & 0.04	\\
3753.0	&	H $_{12}$	&	 0.09  &    &	   &    &    &	   &	   & 0.05	\\
3758.2	&	[\mbox{Fe\,{\sc v}}] 3755.7	&	0.11   &    &	   &    &    &	   &	   & 0.08	\\
3762.4	&	[\mbox{Fe\,{\sc vii}}] 3759.9	&	   &    &	   &    &    &	   &	   & 0.12	\\
3763.4	&	\mbox{O\,{\sc iii}} 3759.9	&	   &    &	   &    &    &	   &	   & 0.21	\\
3773.6	&	H $_{11}$	&	   &    &	   &    &    &	   &	   & 0.08	\\
3800.8	&	H $_{10}$	&	0.15   &    &	   &    &    &	   &	   & 0.11	\\
3838.4	&	H $_9$	&	 0.27  &    &	   &    &    &	   &	   & 0.25	\\
3872.2	&	[\mbox{Ne\,{\sc iii}}] 3868.8	& 0.42   & 0.87   &	   &    &    &	   &	   & 0.07	\\
3892.2	&	H $_8$, \mbox{He\,{\sc i}}	& 0.50   &  1.7  & 0.53   &    &    &	   &	   & 0.30	\\
3930.4	&	\mbox{He\,{\sc i}} 3926.5	&	   &    &	   &    &    &	   &	   & 0.04	\\
3973.2	&	H$_\epsilon$	&	 0.46  &  1.2  & & 0.64   &    &	   &	   & 0.44	\\
4030.0	&	\mbox{He\,{\sc i}} 4026.2	&	0.23   &    &	   &    &    &	   &	   & 0.08	\\
4074.2	&	[\mbox{Fe\,{\sc v}}] 4071.2	&	0.13   &    &	   &    &    &	   &	   & 0.04	\\
4101.4	&	\mbox{N\,{\sc iii}} 4097.3	&	0.16   &    &	   &    &    &	   &	   & 0.14	\\
4105.0	&	H$_\delta$	&  1.1 & 2.1  & 4.0  &  1.1 &  & 1.70 &  &  0.69	\\
4147.8	&	\mbox{He\,{\sc i}} 4143.8	&	   &    &	   &    &    &	   &	   & 0.08	\\
4203.8	&	\mbox{N\,{\sc iii}} 4200.0	&	   &    &	   &    &    &	   &	   & 0.06	\\
4343.8	&	H$_\gamma$	& 1.7  & 3.0  &	6.1  & 1.5  & 1.6  &	2.7  &	1.6  & 	1.2	\\
4367.1	&	[\mbox{O\,{\sc iii}}] 4363.2	& 0.60 &  0.54 &  & 0.25 & & & & 0.06	\\
4392.2	&	\mbox{He\,{\sc i}} 4387.9	&	0.11   &    &	   &    &    &	   &	0.13   & 0.16	\\
4476.0	&	\mbox{He\,{\sc i}} 4471.5	&	0.64   &    &  & 0.32   &    &	   &	0.29   & 0.15	\\
4545.8	&	\mbox{He\,{\sc ii}} 4541.6	&	   &    &	   &    &    &	   &	0.21   & 0.12	\\
4638.6	&	\mbox{N\,{\sc iii}} 4634.2	&	   &    &	   &    &    &	   &	   & 0.08	\\
4643.4	&	[\mbox{Fe\,{\sc ii}}] 4639.7	&	   &    &	   &    &    &	   &	   & 0.05	\\
4645.2	&	\mbox{N\,{\sc iii}} 4640.6	& 0.35$^{\rm a}$   &    &	   &    &    &	   & 0.18$^{\rm a}$   & 0.15	\\
4646.4	&	\mbox{N\,{\sc iii}} 4641.9	&	   &    &	   &    &    &	   &	   & 0.07	\\
4652.0	&	\mbox{C\,{\sc iii}} 4647.4	& 0.28$^{\rm b}$   &    &	   &    &    &	   & 0.08$^{\rm b}$   & 0.12	\\
4654.6	&	\mbox{C\,{\sc iii}} 4650.2	&	   &    &	   &    &    &	   &	   & 0.09	\\
4689.8	&	\mbox{He\,{\sc ii}} 4685.7	&	0.53 & 2.7  & 3.7  & 3.1  &	2.2  &	2.6  &	3.7 &  2.9	\\
4718.0	&	\mbox{He\,{\sc i}} 4613.2	&	0.15   &    &	   &    &    &	   &	   & 0.09	\\
4865.0    &    H$_\beta$    &    4.6  &  9.3  & 18.0  &    5.0  & 6.9  & 8.3  &    4.0 & 2.9    \\
4926.8	&	\mbox{He\,{\sc i}} 4921.9	&	0.23   &    &	   &    &    &	   &	0.26   & 0.24	\\
5012.9	&	[\mbox{O\,{\sc iii}}] 5006.9	&	1.5  &  1.3 & 0.33	 & 0.60 & & 0.16 &	0.15 & 0.09	\\
5020.8	&	\mbox{He\,{\sc i}} 5015.7	&	0.37   &    &	   &    &    &	   &	0.26  & 0.23	\\
5416.7	&	\mbox{He\,{\sc ii}} 5411.5	&	   &    &	   &    &    &	   &	0.24   & 0.24	\\
5881.6	&	\mbox{He\,{\sc i}} 5875.6	&	2.7  & 2.0  &	2.8  &	3.2  &	2.9  &	1.8 &	1.8 & 1.1	\\
6091.8	&	[\mbox{Fe\,{\sc vii}}] 6085.5	&	   &    &	   &    &    &	   &	   & 0.16	\\
6094.4	&	[\mbox{Ca\,{\sc v}}] 6086.4	&	   &    &	   &    &    &	   &	   & 0.15	\\
6567.4	&	H$_\alpha$	& 29.6  &	46.5  &	72.0  &	35.4 &	66.0  &	52.0  &	22.3  &	23.7	\\
6684.5	&	\mbox{He\,{\sc i}} 6678.2	&	1.1	&	1.8  &	2.1  &	1.4  &	2.9  &	2.0  &	0.85 & 0.50	\\
7072.1	&	\mbox{He\,{\sc i}} 7065.2	&	2.6  &	2.1  &	2.6  &	2.2  &	3.0  &	2.0  &	2.0  & 1.4	\\
10132.5	&	\mbox{He\,{\sc ii}} 10123.6	&	   &    &	   &    &    &	   &	   & 0.68	\\
10841.3	&	\mbox{He\,{\sc i}} 10830.3	&	   &    &	   &    &    &	   &	   & 3.3	\\
\hline
\end{tabular}
\raggedright
\textbf{Notes}: $^{\rm a}$~blend of \mbox{N\,{\sc iii}}\,$\lambda\,4640.6$, and $\lambda\,4641.9$; $^{\rm b}$~blend of \mbox{C\,{\sc iii}}\,$\lambda\,4647.4$, and $\lambda\,4650.2$.

\end{table*}

\clearpage

\begin{table}
 \centering
 \begin{minipage}{80mm}
  \caption{UV emission line fluxes in units of 10$^{-13}$ erg cm$^{-2}$ s$^{-1}$.}\label{table_uv}
  \begin{tabular}{|ccccc|}
  \hline

Spectrum	&	swp16591	&	swp25791	&	FOS	&	swp53165	\\
Date	&	21.03.1982	&	27.04.1985	&	19.09.1994	&	23.12.1994	\\
MJD	&	45049	&	46182	&	49614	&	49709	\\
Orbital	phase	&	0.38	&	0.47	&	0.76	&	0.85	\\
\hline
\mbox{N\,{\sc v}} 1240 & 1.6 & 1.3 & 2.8 & \\
\mbox{N\,{\sc iv}]} 1486 & 1.5	& & 1.4 & \\								
\mbox{C\,{\sc iv}} 1554 & 6.0	& &	7.9	&	5.0	\\
\mbox{He\,{\sc ii}} 1640 & 2.4	& 2.9	&	2.4	&	1.6	\\
\mbox{O\,{\sc iii}]} 1663  & 1.5 &  2.8 & 0.8 & \\
\mbox{N\,{\sc iv}} 1720 & 0.6 & & 0.6 & \\
\mbox{N\,{\sc iii}]} 1750 & 1.1 & & 0.6 & \\
\mbox{Si\,{\sc iii}]} 1893 & 0.3 & & 0.1 & \\
\mbox{C\,{\sc iii}]} 1909 & 4.2	& 2.2 &	1.6 &	1.0	\\

\hline
\end{tabular}
\end{minipage}
\end{table}


\begin{table*}
 \centering
  \caption{Photometry of LMC~S63 from the literature.}\label{table_phot}
  \begin{tabular}{ccccccccccc}
  \hline
 \multirow{2}{*}{MJD}	&Orbital & 	\multirow{2}{*}{B}	&	\multirow{2}{*}{V}	&	\multirow{2}{*}{R}	&	\multirow{2}{*}{I} & \multirow{2}{*}{J} & \multirow{2}{*}{H} & \multirow{2}{*}{K} & \multirow{2}{*}{Reference}\\	
 & phase &  & & & &  \cr
	\hline
42421	&	0.89	&	16.08$^J$	&		&	14.62	&	14.05	& &&&	1	\\
44284 & 0.67 && & & & 12.48 & 11.60 &11.33& 2\\
44565$^m$ &  0.93 & 16.3 & & 14.1 & & & & & 3 \\
44934$^m$	&	0.27	&	15.83	&		&	14.29	&	13.61	&	& & &4	\\
45049  & 0.39  & & 15.4 & & & & & & 2\\
45153$^m$	&	0.48	&	15.94	&		&	14.47	&		&	& & &4	\\
46443	&	0.72	&			&		&	14.30$^C$	&	13.86$^C$	&	& & & 5	\\
46768	&	0.03	&	16.91	&	15.98	&		&		&& & &	6	\\
46801	&	0.06	&	16.31	&	15.69	&		&		&& & &	6	\\
46813	&	0.08	&	16.37	&	15.68	&		&		&& & &	6	\\
46827	&	0.09	&	16.28	&	15.65	&		&		&& & &	6	\\
46835	&	0.10	&	16.16	&	15.52	&		&		&& & &	6	\\
46837	&	0.10	&	16.46	&	15.71	&		&		&& & &	6	\\
46845	&	0.11	&	16.19	&	15.62	&		&		&& & &	6	\\
46873	&	0.13	&	16.67	&	15.81	&		&		&& & &	6	\\
46876	&	0.14	&	16.42	&	15.65	&		&		&& & &	6	\\
46877	&	0.14	&	16.19	&	15.54	&		&		&& & &	6	\\
46895	&	0.16	&	16.52	&	15.70	&		&		&& & &	6	\\
46904	&	0.16	&	16.14	&	15.54	&		&		&& & &	6	\\
47127$^m$	&	0.38	&	15.78$^J$	&		&	14.54$^F$	&		&& & &	7	\\
47556	&	0.79	&			&		&14.65$^F$&		&& & &	7	\\
49309$^m$	&	0.47	&	15.96	&	15.31	&		&		&& & &	8	\\
49398 & 0.54 & & & & & 12.54 & 11.66 & 11.33& 9\\
50122	&	0.25	&			&		&		&	13.89	& 12.35 && &	10	\\
50417 & 0.51 && & & 14.00& 12.42 & &11.30& 10\\
50894 & 0.96 &&&& & 12.51 & 11.65 & 11.33& 11\\
51904 & 0.92 &&&&& 12.57 & 11.72 & 11.35 & 12\\
51927 & 0.94 &&&&& 12.54 & 11.66 & 11.31 & 12\\
53332 & 0.28 &&&&& 12.41 & 11.69 & 11.35 & 13\\ 
\hline
\end{tabular}

\raggedright

\textbf{Notes}: $^m$~mean date, $^C$~Kron-Cousins magnitude, $^F$~photographic F magnitude, $^J$~photographic J magnitude\\
\textbf{References}: (1) SuperCOSMOS Sky Surveys \citep{Hamby01}; (2) \citet{kafat83} (3) USNO-A2.0 \citet{Mon98}; (4) The USNO-B1.0 Catalog (Monet, Levine \& Canzian 2003); (5) \cite{Kontizas01}; (6) \cite{lawson90};  (7) The GSC 2.2 Catalogue \citep{STS01}; (8) SPM 4.0 Catalog \citep{Girard11}; (9) \citet{murs96} ; (10) DENIS Catalogue toward Magellanic Clouds \citep{Cioni00}; (11) 2MASS All-Sky Catalog of Point Sources (Cutri et al. 2003); (12) 2MASS 6X Point Source Working Database (Cutri et al. 2006); (13) \citet{kato07}

\end{table*}

\bsp
\label{lastpage}


\begin{thebibliography}{99}

\bibitem[\protect\citeauthoryear{}{}]{}

\bibitem[\protect\citeauthoryear{Alcock et al.}{1991}]{alcock91} Alcock C., Allsman R.A., Alves D.R. et al., 1999, PASP, 111, 1539
\bibitem[\protect\citeauthoryear{Allen}{1980}]{allen80} Allen D.A., 1980, ApL, 20, 131
\bibitem[\protect\citeauthoryear{Angeloni et al.}{2014}]{angel13} Angeloni R., Ferreira Lopes C.E., Masetti N. et al., 2014, MNRAS, 438, 35
\bibitem[\protect\citeauthoryear{Asplund et al.}{2009}]{asplund} Asplund M., Grevesse N., Sauval A., Scott P., 2009, ARAA 47, 481
\bibitem[\protect\citeauthoryear{Bayo et al.}{2008}]{bayo08} Bayo A., Rodrigo C., Barrado y Navascues D., Solano E., Gutierrez R., Morales-Calderon M., Allard F., 2008, A\&A, 492, 277.
\bibitem[\protect\citeauthoryear{Belczy{\'n}ski et al.}{2000}]{belcz00} Belczy{\'n}ski K., Miko{\l}ajewska J., Munari U.,  Ivison R.J., Friedjung M. , 2000, A\&AS, 146, 407
\bibitem[\protect\citeauthoryear{Bergeat et al.}{2002a}]{bergeat2002a} Bergeat J., Knapik A., Rutily B., 2002a, A\&A, 390, 967
\bibitem[\protect\citeauthoryear{Bergeat et al.}{2002b}]{bergeat2002b} Bergeat J., Knapik A., Rutily B., 2002b, A\&A, 390, 987
\bibitem[\protect\citeauthoryear{Bessell et al.}{1983}]{bessell83} Bessell M.S., Wood P.R., Evans T., 1983, MNRAS, 202, 59
\bibitem[\protect\citeauthoryear{Bickert et al.}{1993}]{bickert93} Bickert K.F., Stencel R.E., Luthardt R., 1993, IAUS, 155, 405
\bibitem[\protect\citeauthoryear{Buckley et al.}{2006}]{buc06} Buckley D.A~H., Swart G.P., Meiring J.G., 2006, SPIE, 6267, 32
\bibitem[\protect\citeauthoryear{Burgh et al.}{2003}]{bur03} Burgh E.B., Nordsieck K.H., Kobulnicky H.A, Williams T.B., O'Donoghue D., Smith M.P., Percival J.W., 2003, SPIE, 4841, 1463
\bibitem[\protect\citeauthoryear{Castelli \& Kurucz}{2003}]{cas03} Castelli F., Kurucz R.L., 2003, in IAU Symposium, Vol. 210, Modelling of Stellar Atmospheres, eds. N.E. Piskunov, W.W. Weiss. and D.F. Gray, 2003, ASP-S210
\bibitem[\protect\citeauthoryear{Cioni et al.}{2000}]{Cioni00} Cioni M.R., Loup C., Habing H.J et al., 2000, A\&AS, 144, 235
\bibitem[\protect\citeauthoryear{Cohen}{1979}]{Cohen1979} Cohen M., MNRAS, 186, 837
\bibitem[\protect\citeauthoryear{Crawford et al.}{2010}]{cra10} Crawford S.M., Still M., Schellart P. et al., 2010, Proc. SPIE, 7737E, 54
\bibitem[\protect\citeauthoryear{Clegg}{1987}]{cle87} Clegg R.E.S., 1987, MNRAS, 229, 31
\bibitem[\protect\citeauthoryear{Cutri et al.}{2012}]{cut12}  Cutri R. M., Wright E. L., Conrow T. et al. 2012, VizieR Online Data Catalog, 2311, 0
\bibitem[\protect\citeauthoryear{Feast \& Webster}{1974}]{feast74} Feast M.W., Webster B.L.,  1974, MNRAS, 168, 31
\bibitem[\protect\citeauthoryear{Gaposhkin}{1970}]{gapo70} Gaposhkin S., 1970, SAOSR, 310, 1 
\bibitem[\protect\citeauthoryear{Girard et al.}{2011}]{Girard11} Girard T.M., van Altena W.F., Zacharias N. et al., 2011, AJ, 142, 15
\bibitem[\protect\citeauthoryear{Grevesse \& Sauval}{1998}]{Grev98}  Grevesse N., Sauval A.J., 1998, Space Science Reviews, 85, 161
\bibitem[\protect\citeauthoryear{Gray \& Corbally}{1994}]{Gray94} Gray R.O., Corbally C.J., 1994, AJ, 107, 742
\bibitem[\protect\citeauthoryear{Hambly et al.}{2001}]{Hamby01} Hambly N.C., MacGillivray H.T., Read M.A. et al., 2001, MNRAS, 326, 1279
\bibitem[\protect\citeauthoryear{Iijima}{1981}]{iij81} Iijima T., 1981, in NATO Advanced Study Institute 69, Photometric and Spectroscopic Binary Systems, 517
\bibitem[\protect\citeauthoryear{Kafatos et al.}{1983}]{kafat83} Kafatos M., Michalitsianos A.G., Allen D.A., Stencel R.E, 1983, ApJ, 275, 584
\bibitem[\protect\citeauthoryear{Kato et al.}{2007}]{kato07} Kato D., Nagashima C., Nagayama T. et al., 2007, PASJ 59, 615
\bibitem[\protect\citeauthoryear{Kato et al.}{2012}]{kato2012} Kato M., Miko{\l}ajewska J., Hachisu I., 2012, ApJ, 750, 5
\bibitem[\protect\citeauthoryear{Kenyon et al.}{1991}]{kenyon91} Kenyon S. J., Oliversen N. A., Mikolajewska J., Mikolajewski M., Stencel R.E., Garcia M.R., Anderson C.M., 1991, AJ, 101, 637
\bibitem[\protect\citeauthoryear{Kobulnicky et al.}{2003}]{kobu03} Kobulnicky H.A., Nordsieck K.H., Burgh E.B., Smith M.P., Percival J.W., Williams T.B., O'Donoghue D., 2003, SPIE, 4841, 1634
\bibitem[\protect\citeauthoryear{Kontizas et al.}{2001}]{Kontizas01} Kontizas E., Dapergolas A., Morgan D.H., Kontizas M., 2001, A\&A, 369, 932
\bibitem[\protect\citeauthoryear{Laycock et al.}{2010}]{lay10} Laycock S., Tang S., Grindlay J., Los E., Simcoe R., Mink D., 2010, AJ, 140, 1062
\bibitem[\protect\citeauthoryear{Lawson et al.}{1990}]{lawson90} Lawson W.A., Cottrell P.L., Kilmartin P.M., Gilmore A.C., 1990, MNRAS, 247, 91
\bibitem[\protect\citeauthoryear{Lutz et al.}{2010}]{Lutz2010} Lutz J., Fraser O., McKeever J., Tugaga, D., 2010, PASP, 122, 524
\bibitem[\protect\citeauthoryear{Martin et al.}{2005}]{martin2005} Martin, D. C., Fanson, J., Schiminovich, D. et al., 2005, ApJ, 619, L1
\bibitem[\protect\citeauthoryear{Meier et al.}{1994}]{me94} Meier S.R., Kafatos M., Fahey R.P., Michalitsianos A.G., 1994, ApJS, 94, 183
\bibitem[\protect\citeauthoryear{Miko\l{}ajewska}{2003}]{mikolaj03} Miko\l{}ajewska J., 2003, in Symbiotic Stars Probing Stellar Evolution, eds. R.L.M. Corradi, J. Miko\l{}ajewska, T. Mahoney, ASP Conf. Ser., Vol. 303, 9
\bibitem[\protect\citeauthoryear{Miko\l{}ajewska}{2004}]{mikolaj04} Miko\l{}ajewska J., 2004, RMxAC, 20, 33
\bibitem[\protect\citeauthoryear{Miko\l{}ajewska}{2012}]{mikolaj12} Miko\l{}ajewska J., 2012, BaltA, 21, 5
\bibitem[\protect\citeauthoryear{Miko\l{}ajewska et al.}{1997}]{mikolaj97} Miko\l{}ajewska J., Acker A., Stenholm B., 1997, A\&A, 327, 191
\bibitem[\protect\citeauthoryear{Miko\l{}ajewska et al.}{1995}]{mikolaj95} Miko\l{}ajewska J.,  Kenyon S.J., Mikolajewski M., Garcia M.R., Polidan R.S., 1995, AJ, 109, 1289
\bibitem[\protect\citeauthoryear{Miszalski et al.}{2013}]{misza13} Miszalski B., Miko\l{}ajewska J., Udalski A., 2013, MNRAS, 432, 3186
\bibitem[\protect\citeauthoryear{Miszalski \& Miko\l{}ajewska}{2014}]{mm14} Miszalski B., Miko\l{}ajewska J., MNRAS, 2014, 440, 1410
\bibitem[\protect\citeauthoryear{Miszalski et al.}{2014}]{MMU14} Miszalski B., Miko\l{}ajewska J., Udalski A., 2014, MNRAS, 444, 11
\bibitem[\protect\citeauthoryear{Monet et al.}{1998}]{Mon98} Monet D., Canzian B., Harris H. et al., 1998, USNO-A2.0 (Washington: US Nav. Obs.)
\bibitem[\protect\citeauthoryear{Monet et al.}{2003}]{Mon03} Monet D.G., Levine S.E., Canzian B., 2003, AJ, 125, 984
\bibitem[\protect\citeauthoryear{Morgan}{1992}]{morgan92} Morgan D.H., 1992, MNRAS, 258, 639
\bibitem[\protect\citeauthoryear{Morrissey et al.}{2007}]{morrissey2007} Morrissey P., Conrow T., Barlow T. A. et al., 2007, ApJS,  173, 682
\bibitem[\protect\citeauthoryear{Munari \& Zwitter}{2002}]{munar02} Munari U., Zwitter T., 2002, A\&A, 383, 188
\bibitem[\protect\citeauthoryear{Muerset et al.}{1991}]{murs91} Muerset U., Nussbaumer H., Schmid H.M., Vogel M., 1991, A\&A, 248, 458
\bibitem[\protect\citeauthoryear{Muerset \&  Nussbaumer}{1994}]{murs94} Murset U., Nussbaumer H., 1994, A\&A, 282, 586
\bibitem[\protect\citeauthoryear{Muerset et al.}{1996}]{murs96} Muerset U., Schild H., Vogel M., 1996, A\&A, 307, 516
\bibitem[\protect\citeauthoryear{O'Donoghue et al.}{2006}]{odon06} O'Donoghue D., Buckley D.A.H., Balona L.A. et al., 2006, MNRAS, 372, 151
\bibitem[\protect\citeauthoryear{Oliveira et al.}{2013}]{oliveira13} Oliveira J. M., van Loon J.Th., Sloan G.C. et al., 2013, MNRAS, 428, 3001
\bibitem[\protect\citeauthoryear{Payne-Gaposchkin}{1971}]{payne71} Payne-Gaposchkin C.H., 1971, SCoA, 13, 1
\bibitem[\protect\citeauthoryear{Pereira}{1995}]{pereira95} Pereira C.B., 1995, A\&AS, 111, 471
\bibitem[\protect\citeauthoryear{Percy \& Desjardins}{1996}]{perc96} Percy J.R., Desjardins, A., 1996, PASP, 108, 84
\bibitem[\protect\citeauthoryear{Pietrzy{\'n}ski et al.}{2013}]{pietrz13} Pietrzy{\'n}ski G., Graczyk D., Gieren W. et al., 2013, Nature, 495, 76
\bibitem[\protect\citeauthoryear{Plez}{1998}]{plez98} Plez B., 1998, A\&A 337, 495
\bibitem[\protect\citeauthoryear{Proga et al.}{1994}]{prog94} Proga D.,  Miko\l{}ajewska J., Kenyon S.J., 1994, MNRAS, 268, 213
\bibitem[\protect\citeauthoryear{Rodr\'{i}guez-Flores et al.}{2014}]{rf14} Rodr\'{i}guez-Flores E. R., Corradi R.L.M., Mampaso A., Garc\'{i}a-Alvarez D., Munari U., Greimel R., Rubio-D\'{i}ez M.M., Santander-Garc\'{i}a M., 2014, A\&A, 567, A49
\bibitem[\protect\citeauthoryear{Rodr\'{i}guez-Pascual et al.}{1999}]{rodrig99} Rodr\'{i}guez-Pascual P.M., Gonz\'{a}lez-Riestra R., Schartel N., 1999, A\&AS, 139, 183
\bibitem[\protect\citeauthoryear{Riebel et al.}{2012}]{riebel12} Riebel D., Srinivasan S., Sargent B., Meixner M., 2012, ApJ, 753, 71
\bibitem[\protect\citeauthoryear{Soszynski et al.}{2011}]{soszynski11}  Soszynski I., Udalski A., Szymanski M.K. et al., 2011, AcA, 61, 217
\bibitem[\protect\citeauthoryear{STScI}{2001}]{STS01} Space Telescope Science Institute and Osservatorio Astronomico di Torino, 2001, VizieR Online Data Catalog, 1271, 0
\bibitem[\protect\citeauthoryear{Spano et al.}{2011}]{spano11} Spano M., Mowlavi N., Eyer L., Burki G., Marquette J.B., Lecoeur-Taibi I., Tisserand P., 2011, A\&A, 536, 60
\bibitem[\protect\citeauthoryear{Srinivasan et al.}{2011}]{sriniv11} Srinivasan S., Sargent B.A., Meixner M., 2011, A\&A, 532, 54
\bibitem[\protect\citeauthoryear{Udalski et al.}{2008}]{udalski08}  Udalski A., Szymanski M.K., Soszynski I., Poleski R., 2008, AcA, 58, 69
\bibitem[\protect\citeauthoryear{Van Winckel et al.}{1993}]{win93} Van Winckel H., Duerbeck H.W., Schwarz H.E., 1993, A\&AS, 102, 401
\bibitem[\protect\citeauthoryear{Vernet et al.}{2011}]{vern11} Vernet J., Dekker H., D'Odorico S. et al., 2011, A\&A, 536, 105
\bibitem[\protect\citeauthoryear{Vogel \& Nussbaumer}{1995}]{vogel95} Vogel M., Nussbaumer H., 1995, A\&A, 301, 170
\bibitem[\protect\citeauthoryear{Wood}{2000}]{wood2000} Wood, P., 2000, PASA, 17, 18
\bibitem[\protect\citeauthoryear{Yamashita}{1972}]{Yamashita1972} Yamashita Y., 1972, Ann. Tokyo. Astron. Obs., 13, 169

\bibitem[\protect\citeauthoryear{}{}]{}

\end{thebibliography}
\end{document}